\documentclass{JINST}
\title{Track-based alignment for the ATLAS Inner Detector Tracking System}
\author{Jochen Schieck$^a$ for the ATLAS Collaboration\\
\llap{$^a$}Ludwig-Maximilians-Universit\"at M\"unchen, Am Coulombwall 1, 85748 Garching, Germany and \\
  Excellence Cluster Universe, Boltzmannstr. 2, 85748 Garching, Germany. \\ 
  E-mail: \email{Jochen Schieck@lmu.de}}
\abstract{We discuss the track based alignment of the inner detector of the ATLAS experiment. After describing the main alignment method based on the minimization of  hit residuals, we focus on alignment methods using information from the electromagnetic calorimeter as well as information obtained from physics quantities  such as the invariant mass of the J/$\Psi$-resonance. We  present an overview of the current performance of the ATLAS inner detector and conclude with some general remarks summarizing the experience of the commissioning of the detector from the alignment point of view.}
\begin{document}
\section{Introduction}
The knowledge of the position of sensitive detector elements is crucial for the exploitation of tracking devices. The most precise way to determine this information is by using a track-based alignment algorithm. These algorithms minimize the distance between hits and the corresponding reconstructed track. In this paper, we describe the alignment of the Inner Detector (ID) of the ATLAS experiment~\cite{TheAtlasDetector}, one of the two multi-purpose experiments at the Large Hadron Collider (LHC) at CERN, Geneva. Special emphasis is put on methods to improve the tracking performance using methods beyond the classical residual minimization. \par
\begin{figure}
\begin{center}
\includegraphics[width=.5\textwidth]{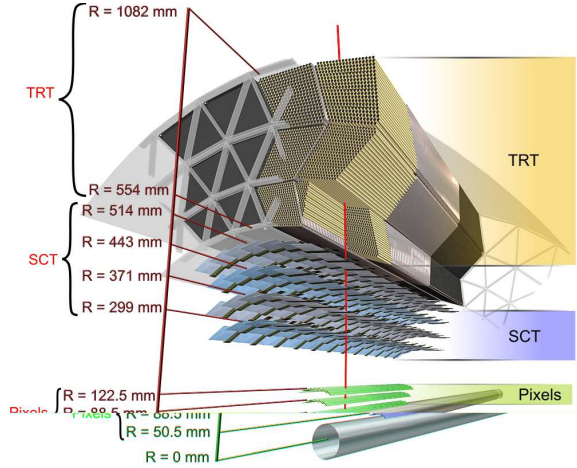} 
\end{center}
\caption{The figure shows the barrel part of ID of the ATLAS experiment with the Pixel, SCT and TRT sub-detectors. }
\label{ATLASID}
\end{figure}

Fig.~\ref{ATLASID} shows the ID of the ATLAS experiment. It consists of three sub-detectors: a pixel detector (Pixels), a silicon strip detector (SCT) and a detector based on drift tubes (TRT). The Pixel and SCT detectors are based on silicon technology and provide very good position resolution below 10 $\mu$m in the Pixel barrel part. \par
To achieve its physics goals, the ATLAS collaboration requires that the resolution of track parameters are degraded by at most 
20$\%$ due to misalignments.
The track parameters affected most are the transverse impact parameter $d_{0}$ and the transverse momentum $p_{T}$, leading to an alignment precision in the most sensitive coordinate in the R$\phi$-plane of 7 $\mu$m for the Pixels and 12~$\mu$m for the SCT~\cite{IDTDR}.
\section{The ATLAS ID track-based alignment algorithm}
The main idea of the track-based alignment consists in the minimization of hit residuals with respect to the corresponding reconstructed tracks. The contribution from hits on all selected tracks are accumulated in a single $\chi^{2}_{0}$-value, which is then minimized with respect to the alignment parameters~\cite{GlobalChi2}. The $\chi^{2}_{0}$ can be written as
\begin{equation}
\chi_{0}^{2}=\sum_{\mathrm{tracks}} [\vec{r}(\vec{a},\vec{\pi})]^{T}V^{-1}[\vec{r}(\vec{a},\vec{\pi})]
\label{trackchi}
\end{equation}
with the sum running over all tracks, $\vec{r}(\vec{a},\vec{\pi})$ being the hit residuals being dependent on the alignment parameters $\vec{a}$ as well as the track parameters $\vec{\pi}$, and $V$ being the the covariance matrix. Correlations between the different detector elements (between modules as well as between sub-detectors) are taken into account by using the track parameter dependence of the hit residuals, leading to a nested dependence of the derivative:
\begin{equation}
\frac{d\vec{r}}{d\vec{a}}=\frac{\partial\vec{r}}{\partial\vec{a}} + \frac{\partial\vec{r}}{\partial\vec{\pi}}  \cdot \frac{d\vec{\pi}}{d\vec{a}}  
\end{equation}  
The $\chi^{2}_{0}$ is then taylor-developed around the actual alignment position in order to estimate the minimal $\chi^{2}$-value. The calculation returns $\vec{a}$ which represents the required detector movement in order to bring the overall $\chi^{2}$ to its minimum. This requirement is necessary but not sufficient to solve the alignment problem. The solution could return a local $\chi^{2}$-value only because of only weakly determined movements, the so called weak modes. \par
\begin{figure}
\begin{tabular}{cc}
\includegraphics[width=.5\textwidth]{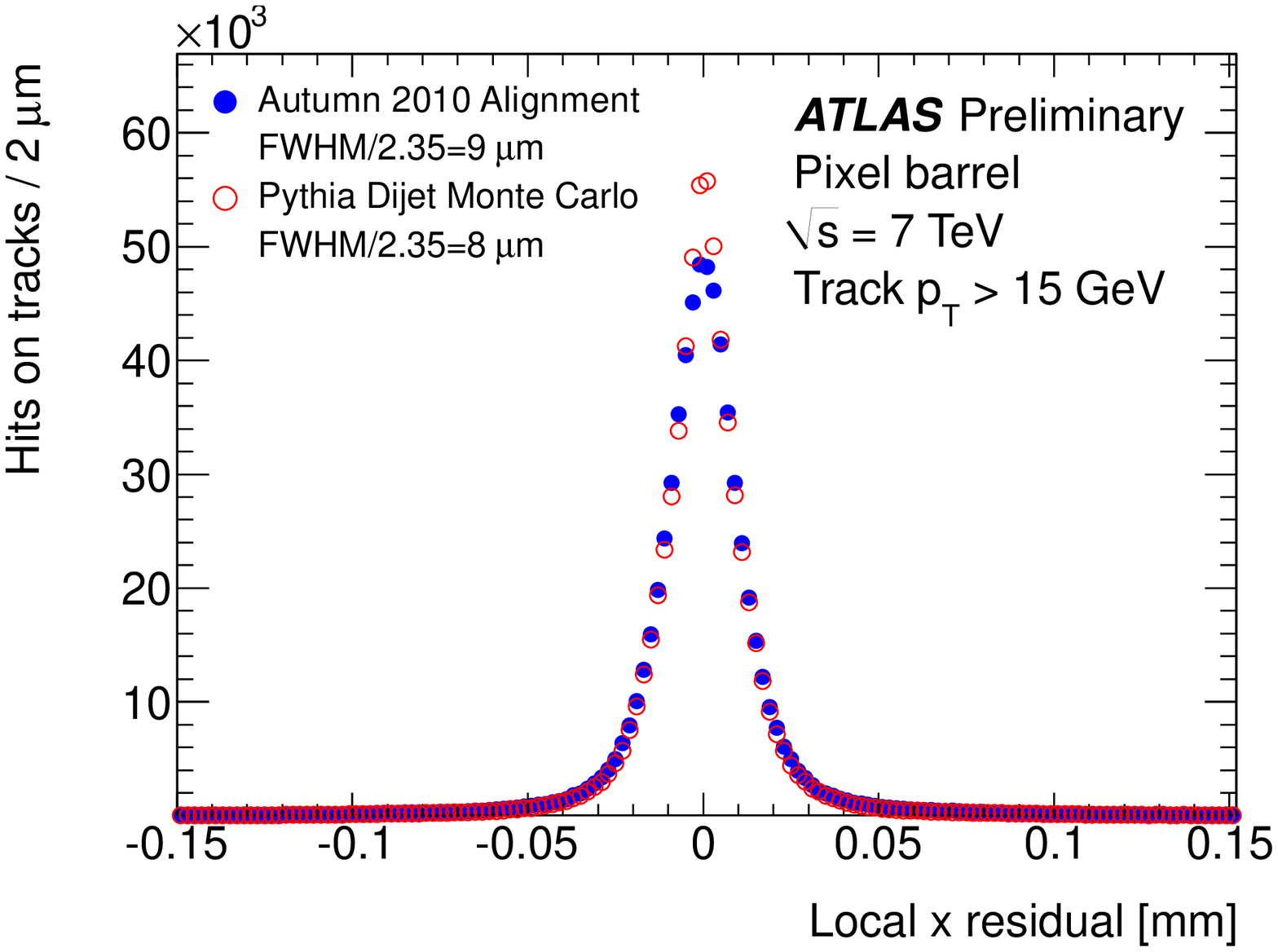} & 
\includegraphics[width=.5\textwidth]{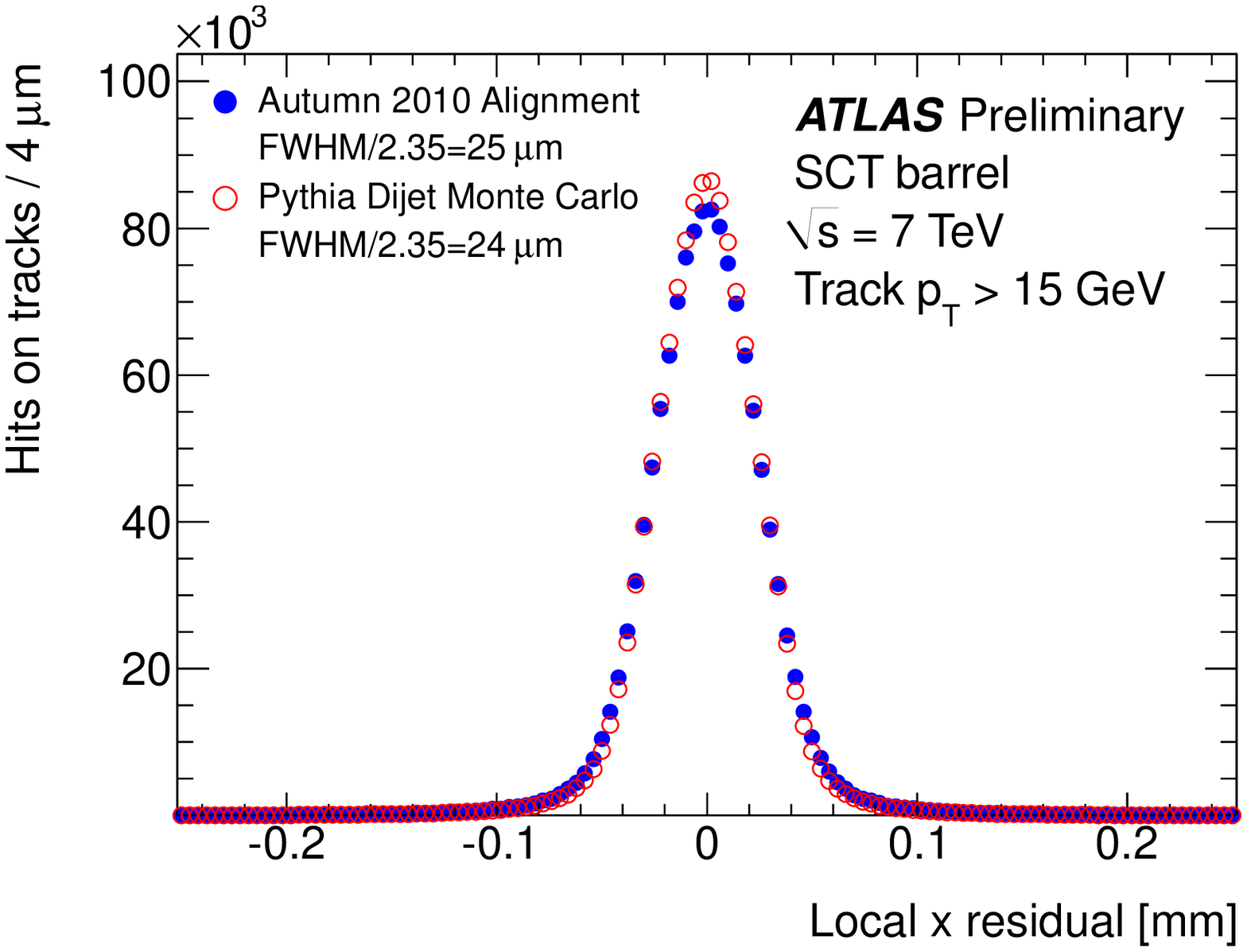} \\
\end{tabular}
\caption{The figures represent the residual distribution compared to the expected residual distribution obtained from simulated events with perfect geometry. The left plot shows the residual distribution using Pixel barrel hits only, while the right one shows SCT barrel hits only.}
\label{FigResidual}
\end{figure}

This alignment step is repeated several times following the hierarchical structure of the detector. First the large structures, like the barrel with respect to the end caps are aligned. At the very end all individual modules are aligned. It should be also not noted, that not all degrees of freedom are aligned for all individual alignment steps. For the track based alignment we use tracks with two different topologies: tracks with high momentum originating from the interaction point as well as tracks reconstructed from cosmic ray showers. The residual distribution obtained after this alignment step almost matches the expected distribution obtained from simulated events with perfect geometry, as shown in Figure~\ref{FigResidual}~\cite{Alignment}.
\section{External alignment constraints}
Besides the minimization of hit residuals, we use additional information to improve the knowledge of the detector position. During the first year of data taking, the main focus was put on improving the understanding of the systematic biases on the  momentum measurement. For this we use the dependence of the invariant mass on the azimuthal angle $\phi$~\footnote{The coordinate system of ATLAS is a right-handed coordinate system with the x-axis pointing towards the centre of the LHC ring, and the z-axis along the tunnel. The y-axis is slightly tilted with respect to vertical due to the general tilt of the tunnel. The angle $\phi$ is defined in the $xy$-plane.} as well as the $E/p$ information obtained from a comparison of the ID momentum measurement with the electromagnetic calorimeter energy measurement.
\subsection{Tilt of the solenoidal field with respect of the ID}
The track-based alignment of the ID provides only an internally consistent alignment. Six degrees of freedom, representing the overall position of the detector in space cannot be determined. After the production of the first set of alignment constants, using tracks from cosmic rays only, it was decided to maintain the overall position of the ID. We require the $\chi^{2}$ of all distances between the module position from the previous set of alignment constant and the newly determined set of alignment constants to be minimal. The beam-spot position is preserved by requiring for all sets of alignment constants  the tracks to originate in the transverse plane from the previously determined beamspot.\par
The first set of alignment constants did not take any information from the solenoidal field into account. The direction of the magnetic field with respect to the ID is only correct within the placement precision of the ID within the cryostat containing the solenoid. However, the direction of the magnetic field is crucial for the measurement of the track momentum and a rotation of the ID with respect to the origin will lead to distorted measurements of the transverse momentum $p_{T}$. Due to the larger lever-arm, a misalignment will be most prominent in the end-caps of the detector. \par
\begin{figure}
\begin{tabular}{cc}
\includegraphics[width=.5\textwidth]{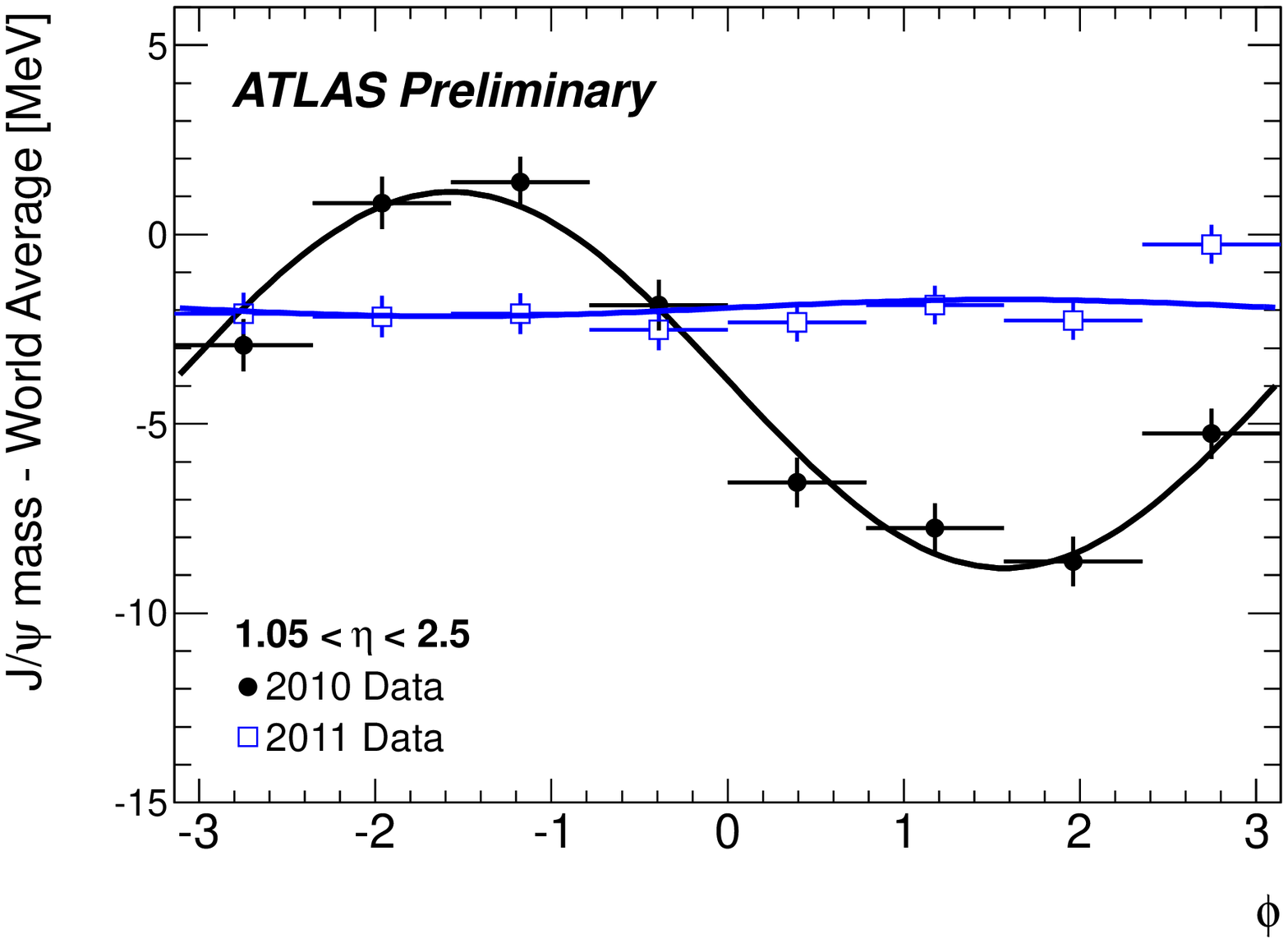} & 
\includegraphics[width=.5\textwidth]{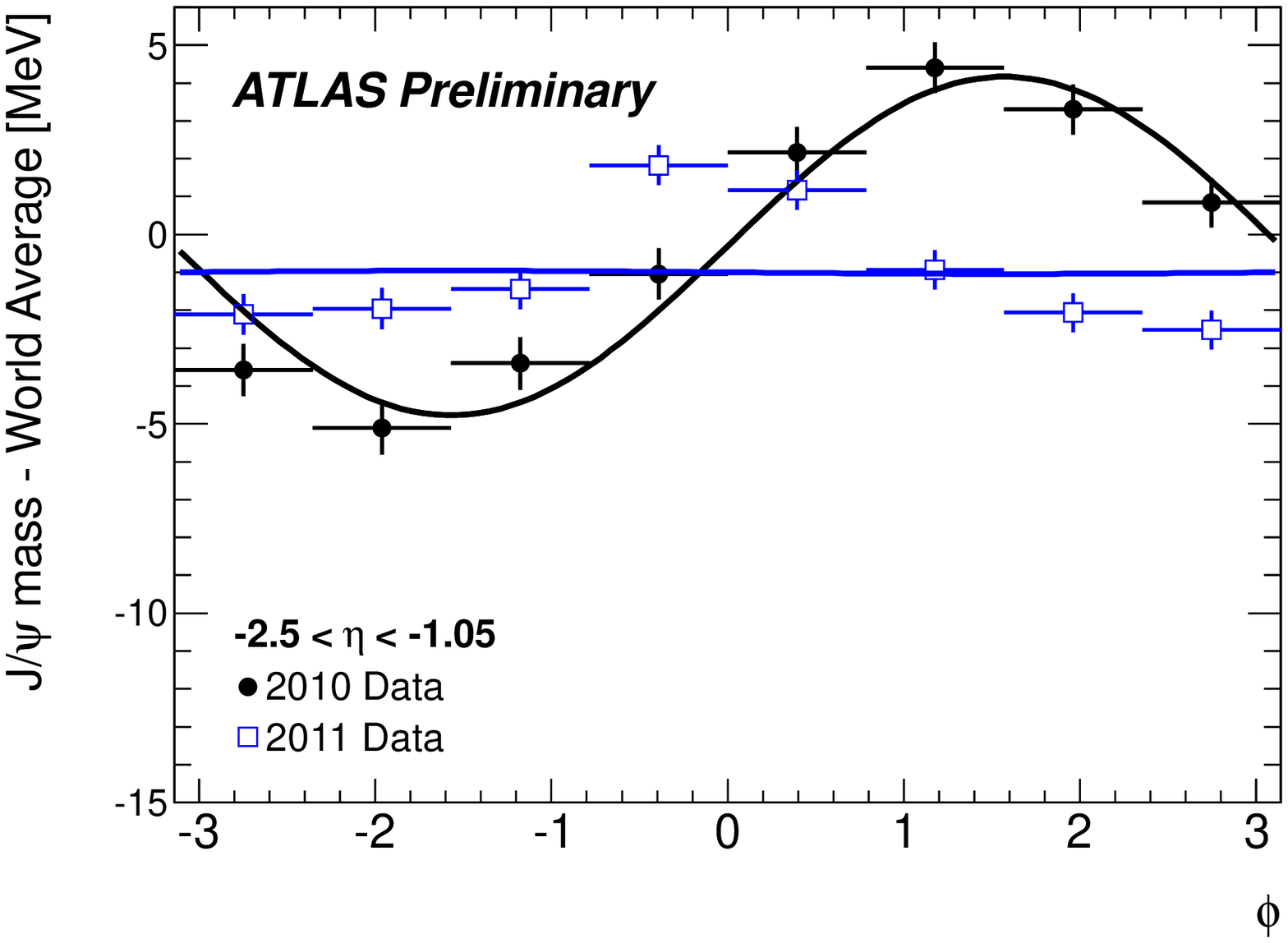} \\
\end{tabular}
\caption{The figures show the invariant J/$\psi$-mass as a function of the azimuthal angle $\phi$ for both end-caps. The left plot corresponds to the end-cap at positive $Z$, while the right plot corresponds to the end-cap at negative $Z$. The black points are obtained with a set of alignment constants prior to the rotation correction around the global x axis. The line is obtained by a sinusoidal fit to the data point. The blue points represent the J/$\psi$-mass as a function of $\phi$ after the rotation correction of 55 mrad.  }
\label{FigTilt}
\end{figure}

To quantify this systematic effect and in order to correct for this we measure the invariant mass of the J/$\psi$-resonance as a function of the azimuthal angle $\phi$ in the two end-cap regions of the detector. A clear $\phi$-dependence of the invariant J/$\psi$-mass is visible in both end-caps, originating from the tilt of the magnetic field. This distortion can be explained by a rotation of the ID around the x-axis  by 55 mrad. The impact of this rotation is shown in Fig.~\ref{FigTilt}. The strong $\phi$-dependence of the invariant J/$\psi$-mass almost disappeared after the rotation correction.
\subsection{Improved momentum scale using information from the electromagnetic calorimeter}
\begin{figure}
\begin{center}
\includegraphics[width=.70\textwidth]{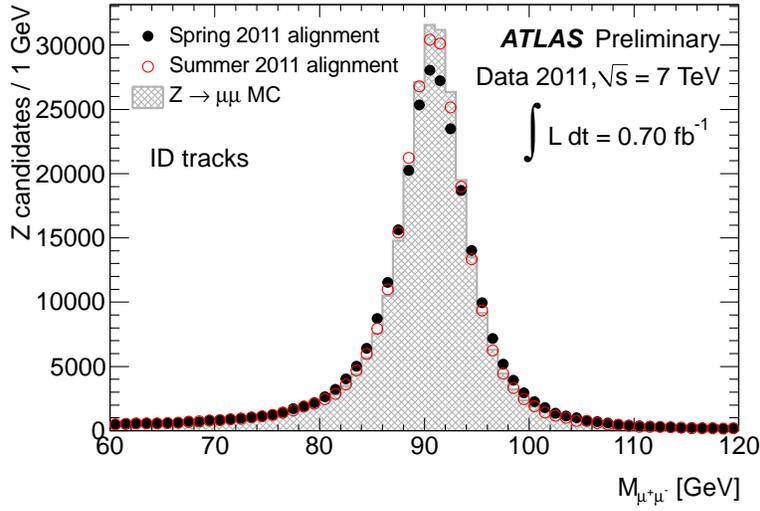} 
\end{center}
\caption{The figure shows the improved width of the invariant mass distribution of the $Z$-boson determined with a set of alignment constants requesting identical $\frac{E}{p}$-response for positively and negatively charged electrons. The distribution indicated by the black points are obtained with a set of alignment constants without $\frac{E}{p}$-constraints, while the distribution with red points is obtained with a set of alignment constants with the $\frac{E}{p}$-constraint implemented. The grey shaded area shows the expectation from simulated events using perfect geometry.}
\label{ZInvMass}
\end{figure}

To further improve the momentum measurement, the electron energy measurement in the
electromagnetic calorimeter can be used in the alignment procedure~\cite{TheBigPawel}. The sagitta distortion~\footnote{Sagitta distortions refer to systematic distortions of the geometry which transform helical into helical tracks, but bias the track parameters systematically.}
leads to a charge asymmetric bias in the measurements which can be generically expressed by:
\begin{equation}
\frac{Q}{p^{\mathrm{correct}}}=\frac{Q}{p^{\mathrm reconstructed}}(1 + Q \cdot p_{T}^{\mathrm reconstructed}\cdot \delta),
\end{equation}
where $\delta$ describes the local sagitta bias parameter and $Q$ the particle charge. In order to determine
the sagitta bias we use the fact that the energy response of the electromagnetic calorimeter to electrons and positrons is independent of the charge. The scaling factor $\delta$ can be determined from the average ratio $<E/p>$ measured for electrons and positrons:
 \begin{equation}
\left< \frac{E}{p^{\mathrm{correct}}} \right>^{\pm} =\left< \frac{E}{p^{\mathrm{reconstructed}}} \right>^{\pm}  \left( 1+ Q \cdot \left<  p_{T}^{\mathrm{reconstructed}}\right>^{\pm} \cdot \delta \right)
\end{equation}
and negatively charged tracks
\begin{equation}
\left< \frac{E}{p^{\mathrm{correct}}} \right>^{-} =\left< \frac{E}{p^{\mathrm{reconstructed}}} \right>^{+} \left( 1- \left<  p_{T}^{\mathrm{reconstructed}}\right>^{-} \cdot \delta \right),
\end{equation}
where $<\,>$ indicates an average over all measurements. By requesting the average ratio to be identical for negatively and positively charged tracks the scaling factor $\delta$ is determined. The improved performance of the ID using reconstructed tracks is demonstrated using a measurement of the width of the invariant mass of the $Z$-boson, obtained with a set of alignment constants biased to return identical $\frac{E}{p}$-response for e$^{+}$ and e$^{-}$, is shown in Fig.~\ref{ZInvMass}. The width of the $Z$-boson measured with tracks reconstructed in the barrel region of the Inner Detector decreases from $1.87\pm0.02$ GeV/$c^{2}$ (Spring 2011)  to $1.75\pm0.02$ GeV/$c^{2}$ (Summer 2011), compared to  $1.60\pm0.01$ GeV/$c^{2}$ obtained with simulated events.
\section{Measurement of the impact parameter} 
During the alignment procedure all tracks used in Eq.~\ref{trackchi} are constrained to originate from the beam-spot region.
This corresponds  to the track impact parameter ($d_{0}$) distribution being centered around zero. 
However, a measurement of the impact parameter as a function of $\phi$ and $\eta$ returned in some regions systematic biases in $d_{0}$ of up to 10 $\mu$m. This distorted measurement might lead to some systematic uncertainties in the measurement of the lifetime-dependent measurements, such as the measurement of the $B$-meson lifetime or tagging of $B$-hadrons using impact parameter based observables. 
\section{Time dependence of the ATLAS ID alignment}
Currently one set of alignment constants is used for the complete 2011 datataking period. Based on this set of alignment constants an additional alignment is performed with the Pixels, the SCT barrel and end-caps and the TRT barrel and end-caps treated as rigid bodies. The alignment algorithm returns small translations along the x-axis for the SCT and the TRT of up to 10 $\mu$m. The position of the pixel detector, treated as one solid object, is stable with respect to the beam spot position during the studied period. The observed translations can be related to sudden environmental changes, such as failures of the cooling system, power cuts or a ramp down of the muon spectrometer toroid.
\section{Final remarks and Conclusion}
Before concluding this article, we would like to summarize some important observations made during the commissioning of the ATLAS ID alignment algorithm:
\begin{itemize}
\item For the production as well as for the validation and cross-check of the alignment constants, the same data is processed several times, including track finding and track fitting. This is by far the most time consuming part in the overall alignment procedure. Track-based alignment imposes strong requirements on the selection of tracks, such as setting high $p_{T}$ thresholds in order to minimize multiple-scattering effects. With  these stringent cuts, only a reduced set of tracks will be finally used in the alignment procedure. Introducing a reduced data format or set, which contains only the information required for the track-based alignment (i.e. only hits associated to high $p_{T}$-tracks) the turnaround of the alignment production will improve considerably. Including hits in a wide road around the initially reconstructed track will prevent problems in cases of large initial misalignments and opens the opportunity to select different hits for different iterations of track finding. 
It was found that this reduced data format improves the production and the validation of the alignment considerably.
\item The trigger during collisions is optimized to select tracks originating from the interaction point and events with tracks from cosmic ray events are rejected. For the track-based alignment of the ATLAS ID, tracks from cosmic rays events are of great importance, since they relate different parts of the detector together, i.e. the top and the bottom part of the detector. Determining alignment constants can be seen as inverting a matrix built as described by Eq.~\ref{trackchi}. Tracks with different track topologies, compared to tracks originating from the origin, populate off-diagonal elements of the matrix, leading  to an improved condition of the matrix.
\item The minimization of hit residuals as described by Eq.~\ref{trackchi} is necessary for determining the correct position, but not sufficient. The algorithms used by the tracks based alignment center the residual distribution around zero, and for this reason the mean value of the residual distribution is not a good test to check the quality of the alignment. Similarly the width of the residual distribution can be similar to the expected distribution obtained with perfect geometry without being perfectly aligned. The shape of the residual distribution reflects only  the correct implementation of the track-based alignment algorithm. To draw further conclusions about the quality of the alignment, additional observables, not based on hit residuals, are required. Typical observables are invariant mass distributions and their variation with respect to different regions of the detector. It is essential to focus on non-residual based observables at a very early stage and not get misled by centered residual distributions. 
\end{itemize}
\section{Conclusions}
In this article, we discuss the track-based alignment of the Inner Detector of the ATLAS experiment. After briefly explaining the main principles of track-based alignment, we focus on improving the tracking performance by using additional information besides the standard hit-residual minimization. We present a method based on measuring the $\phi$-variation of invariant mass distributions to correct for a possible tilt of the position-sensitive detectors with respect to the solenoidal field. A second technique uses the fact that the response of positively charged and negatively charged electrons in the calorimeter is identical while the momentum could deviate significantly due to residual misalignment effects. Both approaches improve the the tracking performance significantly, resulting in  the ATLAS Inner Detector 
approaching closely its design performance already one year after first data taking.
\acknowledgments
This research was supported by the DFG cluster of excellence "Origin and Structure of the Universe" (www.universe-cluster.de).


\begin{thebibliography}{9}
\bibitem{TheAtlasDetector}
The ATLAS Collabortation \emph{The ATLAS Experiment at the CERN Large Hadron Collider}, JINST, 3, S08003, 2008.

\bibitem{IDTDR}
S.~Haywood et al., \emph{The ATLAS Inner Detector Technical Design Report}, CERN/LHCC/97-16.

\bibitem{GlobalChi2}
P.~Br\"uckman et al., \emph{Global $\chi^{2}$ approach to the alignment of the ATLAS silicon tracking detectors}, ATLAS Note ATL-INDET-PUB-2005-002.

\bibitem{Alignment}
The ATLAS Collaboration, \emph{Alignment of the ATLAS Inner Detector Tracking System with 2010 LHC proton-proton collisions at $\sqrt{s}$ = 7 TeV}, ATLAS-CONF-2011-012.

\bibitem{TheBigPawel}
P.~Br\"uckman (for the ATLAS Collaboration),\emph{ Alignment of the ATLAS Inner Detector Tracking System - Solving the Problem}, N.~Phys.~B (Proceedings Supplements)  197 (2009) 158.

\bibitem{Btag}
The ATLAS Collaboration, \emph{Commissioning of the ATLAS high-performance $b$-tagging algorithms in the 7 TeV collision data}, ATLAS-CONF-2011-102.


\end{thebibliography}
\end{document}